\documentclass[secnumarabic,graphics,floatfix,nofootinbib,
tightenlines,nobibnotes,aps,prl,12pt]{revtex4-1}
\usepackage{amsmath}  % needed for \tfrac, \bmatrix, etc.
\usepackage{amsfonts} % needed for bold Greek, Fraktur, and blackboard bold
\usepackage{graphicx} % needed for figures

\begin{document}

% Be sure to use the \title, \author, \affiliation, and \abstract macros
% to format your title page.  Don't use lower-level macros to  manually
% adjust the fonts and centering.

\title{Complex Faraday and Kerr Rotations in Right and Left Handed Films}
% In a long title you can use \\ to force a line break at a certain location.

\author{Josh Lofy}
\email{JLofy1@csub.edu}
\author{Vladimir Gasparian}
\affiliation{Department of Physics, California State University, Bakersfield, CA 93311, USA}
\author{Zhyrair Gevorkian}
\affiliation{Yerevan Physics Institute, Yerevan, Armenia}
\affiliation{Institute of Radiophysics and Electronics, Ashtarak-2, 0203, Armenia}

\date{\today}

\begin{abstract}
By studying the rotations of the polarization of  light propagating in right and left handed films, with emphasis on the transmission (Faraday effect) and reflections (Kerr effect) of light and through the use of complex values representing the rotations, it can be shown that the real portions of the complex angle of Faraday and Kerr rotations are odd functions with respect to the refractive index n and that the respective imaginary portions of the angles are an even function of n.  Multiple reflections within the medium lead to the maximums of the real portions of Faraday and Kerr effects to not coincide with zero ellipticity.  It will also be shown that in the thin film case with left handed materials there are large resonant enhancements of the reflected Kerr angle that could be obtained experimentally.
%The angles of rotation for the respective effects have some maximums near when the ellipticity of the reflections are zero.  As well, with left handed ultra thin films under specific circumstances it can be shown that a large resonant enhancement of the reflected Kerr angle can also be obtained.  Focus on ellipticity as zero does not mean maximum of real potion
\end{abstract}
% AJP requires an abstract for all regular article submissions.
% Abstracts are optional for submissions to the "Notes and Discussions" section.

\maketitle % title page is now complete

\section{Introduction} % Section titles are automatically converted to all-caps.
% Section numbering is automatic.

Negative refractive index magneto optical metamaterials (also called left handed materials (LHM)) are a new type of artificial material characterized by having both permittivity $\epsilon$ and permeability $\mu$ negative \cite{VE68,SM00,NIMM}. Despite the fact that even with $\epsilon <0$ and $\mu <0$, these metamaterials do have negative refractive index ($n=\sqrt{\epsilon\mu}$). LHM have multiple uses: they may be used to resolve images beyond the diffraction limit \cite{PE00,SM04}, act as an electromagnetic cloaks for particular frequencies of light \cite{L2006,S2006,MMCLOAK}, enhance the quantum interference \cite{Y} or yield to slow light propagation \cite{P}. The presence of negative indices of refraction in one-dimensional (1D) disordered metamaterials strongly suppresses Anderson localization, due to the lack of phase accumulation during wave propagation which thus weakens interference effects necessary for localization \cite{AS07}. As a consequence, an unusual behavior of the localization length $\xi$ at long-wavelengths $\lambda$ has been observed \cite{AS07,BA12,G15}. This is unlike the well-known quadratic asymptotic behavior $\xi \sim \lambda^2$ for standard isotropic layers (see, e.g. \cite{T13}). It can be seen that the metamaterial configurations have an affect on the magneto-optical transport properties of the electromagnetic waves.

 Particularly, the sign of plane polarization rotation angle in a left handed medium (LHM) is opposite to the sign of rotation angle in a right handed medium (RHM).
  The Faraday and Kerr rotations (FR \& KR) are non-reciprocal polarization rotation effects in that the sign of the rotation is always relative to the direction of the magnetic field. This is different in optically active media rotate where the rotation of the polarization is relative to the direction of the wave vector. Thus, the non-reciprocity of the Faraday and Kerr effects allow light to accumulate rotations of the same sign and magnitude for both forward and backward propagation and can be enhanced even further by additional round-trip reflections through the medium.

If we assume no absorption and neglect the influence of the boundaries  of the system, then in bulk materials the Faraday rotation angle is at a maximum for a given $\epsilon$, $\mu$, Verdet constant, and length of medium L that light travels within the medium for a constant linear magnetic field. When the reflection within the boundaries is important, the outgoing reflected wave is generally elliptically polarized even without absorption, where the major axis of the ellipse is rotated with respect to the original direction of polarization and the maximum FR (KR) angle does not necessarily coincide with angular frequencies $\omega$ of light at which zero ellipticity can be measured (we will come back to this question in section II).

  The real part of the rotation angle describes the change of polarization in linearly polarized light. The imaginary part describes the ellipticity of transmitted or reflected light.
  Once we know the scattering matrix elements $r$ and $t$ of a one-dimensional light propagation problem, then the two characteristic parameters of Faraday/Kerr rotation (Real) and Faraday/Kerr ellipticity (Imaginary) of the magneto-optical transmission/reflection measurements can be written in complex form as the real and imaginary parts of a well-defined complex angle $\theta^T$ and $\theta^R$ (see Eqs. (\ref{FR1}) and (\ref{FR2}).

In the present paper, we theoretically consider the Faraday rotation of light passing through a RHM/LHM  film of thickness L taking into account the multiple reflections from the boundaries. This exactly solvable simple model is chosen on purpose to present different aspects of RHM and LHM. It will be shown that the real part of the complex angle of Faraday rotation is an odd function with respect to the refractive index n, while the imaginary part of the angle is an even function of n. We have obtained the rotation angle of backscattered light (Kerr effect) from the RHM/LHM film as well. In the limit of ultra thin LHM film under specific circumstances we will see a large resonant enhancement of the reflected KR angle.

The work is organized as follows. In section II we formulate the problem with appropriate analytical expressions for the complex Faraday angle of transmitted light. In section III we analyze the Kerr effect and calculate the real and imaginary angles of reflection.

\section{Right handed and Left handed dielectric slab}

Let us consider a slab: confined to the segment $0\le x \le L$, with a positive surface impedance $z=\sqrt{\mu/\epsilon}$ for either RHM or LHM, and characterized by permittivity $\epsilon =n/z$ and permeability $\mu=nz$. Both n and z, and therefore $\epsilon$ and $\mu$, are frequency dependent complex functions that satisfy certain requirements based on causality. For passive materials, {\it Re(z)} and {\it Im(n)} must be greater than zero.

The two semi-infinite media outside of the slab are the same and are characterized by the dielectric constant $\epsilon_1$. A linearly polarized electromagnetic plane wave with $\omega$ angular frequency enters the slab from the left at normal incidence. We take the direction of propagation as the x axis, and that of the electric field ${\vec E_0}$ in the incident wave as the z axis. A weak magnetic field $\vec B$, is applied in the x direction and confined to the slab which causes the direction of linear polarization to rotate as light propagates through the medium. As a consequence, the dielectric tensor develops non-zero off-diagonal elements. Magneto-optic effects  are  related  to  the  off-diagonal  component $\epsilon_{ij}$  ($i,j \in\{1,2\}$) ,  whereas  optical  properties  are related  to  the  diagonal  component $\epsilon_{ii}$ . The magnitude  of  the  off-diagonal  component $\epsilon_{ij}$  is  two  orders  of  magnitude  smaller  than  that  of the  diagonal  component $\epsilon_{ii}$. The generalized principle of symmetry of kinetic coefficinets implies that $\epsilon_{ij}(\vec B)=\epsilon_{ji}^*(-\vec B)$. The condition that absorption is absent requires that the tensor should be Hermitian $\epsilon_{ij}=\epsilon^{*}_{ji}$: the  diagonal  components  of  the  dielectric  tensor  are  even  functions  of  an  applied  magnetic  field,  and  the  off-diagonal  components  $\epsilon_{ij}$  are odd functions and have first-order  magnetic  field  dependence. The dielectric tensor of the slab is given by \cite{LL}
\begin{equation}
\epsilon_{ij}=\left( \begin{array}{cc}
\epsilon & +ig\\
-ig & \epsilon\\
\end{array}
\right),\label{eps}
\end{equation}
where  $\vec g$ is the gyration vector directed on the magnetic-field direction.  We absorb the external magnetic field $\vec{B}$ into the gyrotropic vector $g$ for our $\epsilon_{ij}$ to make our calculations valid  for the cases of external magnetic fields and magneto-optic materials.

The components $E_z$ and $E_y$, $H_z$ and $H_y$ in the film are not constant, where these values depend only upon the coordinate x. As well, when a magnetic field is applied in the x -direction, the off-diagonal elements $\epsilon_{ij}$ cause coupling between the  $E_z$ and $E_y$ electric field ($H_z$ and $H_y$) components. The linearly polarized incident electromagnetic wave now can be presented as the sum of circularly polarized waves with opposite directions of rotation, which propagate through the slab with a different wave vector $k_{\pm}=\omega n_{\pm}/c$.  For circularly polarized waves $E_{\pm}=E_y\pm iE_z$ the Maxwell equations have the form \cite{LL}:
\begin{equation}
\frac{\partial^2 E_{\pm}}{\partial x^2}+\frac{\omega^2\epsilon_{\pm}}{c^2}E_{\pm}=0,
\end{equation}
where $\epsilon_{\pm}=\epsilon\pm g$.

The reflectance and transmittance amplitudes can be obtained using the continuity of the tangential components of the electric (magnetic) fields at the two interfaces, $x=0$ and $x=L$. Solving the equations with the appropriate boundary conditions at $x=0$ and $x=L$ we obtain for the transmitted waves
 $E^{\prime}_{+}$ and $E^{\prime}_{-}$
$$E^{\prime}_{\pm} =  E_{0}t_{\pm},$$
where $t_{\pm}$ is the transmission amplitude for right and left circularly polarized light and can be presented in the form \cite{LL}:

\begin{equation}
t_{\pm}= T^{1/2}_{\pm}{e^{-i\omega Ln_{\pm}/c}e^{i\psi_{\pm}}}. \label{ta}
\end{equation}

The coefficient of transmission $T_{\pm}$ and the phase $\tan{\psi}_{\pm}$ are given by the following expressions, respectively
\begin{equation}
T_{\pm}=\bigg[{{1+\frac{1}{4}\bigg(z_{\pm}-\frac{1}{z_{\pm}}\bigg)^2\sin^2(\omega Ln_{\pm}/c)}}\bigg]^{-1}, \label{T}
\end{equation}
\begin{equation}
\tan{\psi}_{\pm}=\frac{1}{2}\bigg(z_{\pm}+\frac{1}{z_{\pm}}\bigg)\tan({\omega Ln_{\pm}/c}).\label{p}
\end{equation}
It can been proven very generally that there is a linear relation between the real and imaginary parts of $t_{\pm}$ or between $\ln t_{\pm}$ and ${\psi}_{\pm}$. These well known linear Kramers-Kronig relations can be rewritten in terms of localization length and density of states \cite{thou}. The complex FR angle with the imaginary and real parts is introduced (see, e.g., \cite{1995})
\begin{equation}
\theta^{T} =-\frac{i}{2}\ln\frac{t_{+}}{t_{-}}=\frac{\psi_{+}-\psi_{-}}{2}-\frac{i}{2}\ln\frac{T^{1/2}_{+}}{T^{1/2}_{-}}\equiv
\theta_1^{T}+i\theta_2^{T}.\label {FR1}
\end{equation}

As is seen from Eq. (\ref{FR1}), if $T_{+} = T_{-}$, then $\theta^{T}\equiv \theta_1^{T}$ would be real; this signifies that the wave remains linearly polarized with vector $\vec E$ rotated through the angle $\theta^{T}$ to the initial direction. In the Faraday geometry, or when a  magnetic  field  is  applied parallel  to  the direction of light  propagation, and in the absence of material losses within a thin film ($R+T=1$, where R is the reflection coefficient), $T_{+} = T_{-}$ if: (i) the sample is infinite (no boundaries), (ii) for certain thicknesses  total transmission occurs, that is $T=1$ and (iii) $n\frac{\partial T^{1/2}}{\partial n}=z\frac{\partial T^{1/2}}{\partial z}.$ The third condition implies that at a certain thicknesses $\theta_2^{T}$ becomes zero (the solutions of the following transcendental equation, $x_0=\frac{z^2+1}{z^2-1}\tan x_0$). At these points the transmission coefficient T, in contrast to the two previous cases, is not one and its value decreases with increasing $x_0$ with a saturated value of $4z^2/(z^2+1)^2$ for $x_0$ tends $\infty$. This saturated value corresponds exactly to one-quarter wavelength.   In the case of a more complex geometry (for example a multi-layered periodic structure in an external magnetic field) it is possible for the Faraday rotation $\theta^{T}$ to be real more than three times with a simultaneously zero imaginary portion. If $T_{+}\ne T_{-}$, the light has an elliptical polarization and is not simply linearly polarized.  The ratio of the ellipses semi-axis is determined by relation ($b < a$)
\begin{equation}
\frac{b}{a} = |\tan \theta_2^{T}|=\frac{\bigg|T^{1/2}_+ -T^{1/2}_-\bigg| }{\bigg|T^{1/2}_+ +T^{1/2}_-\bigg|}, \label{ration}
\end{equation}
and with an angle between the large axis of the ellipse and the $y$ axis, as
\begin{equation}
\theta_1^{T} =\frac{\psi_{+}-\psi_{-}}{2}.
\end{equation}
For bulk (isotropic) samples or optical devices, where one-way light propagation is important, ${\psi}_{\pm}={\omega Ln_{\pm}/c}$ (see Eq. (\ref{p}) when $z_{\pm}=1$) the result reads
\begin{equation}
\theta_1^{T} =\frac{L\omega \sqrt\mu}{2c}\bigg(\sqrt{\epsilon+g}-\sqrt{\epsilon -g}\bigg).\label{theta}
\end{equation}
Increasing of the linear polarizations rotation $\theta_1^{T}$ in a small length scale can be done in many different ways: (i) by taking into account the multiple reflections which in a finite layer or in resonant structures can lead to an enhancement of the FR angle in comparison with a single direct pass (for example Fabry-Perot cavities filled with a magneto-optic material \cite{FP}), (ii) tuning the optical properties of permittivity $\epsilon$, $\mu$ and $g$ by the modification of the structure size and shape of the material; varying the composition of alloyed and intermetallic nanostructures.  (iii) using metamaterials to tailor the optical properties of the host system \cite{met1}
 (iv) change the dielectric permittivity tensor of a medium with time, etc. The time dependence case may concern both to diagonal and non-diagonal permittivity terms of $\epsilon_{ij}$ (see Eq. (\ref{eps})).

In Ref. \cite{met1} the permittivity tensor of a magneto-optical material is tailored by embedded wire meshes. These wires can only tune the diagonal element of the permittivity tensor in terms of topological parameters and material properties and thus, effectively reducing $\epsilon$ to the value of $g$ (creating a near zero epsilon (NZE) metamaterial) \cite{NZE}. For such frequencies the second term of Eq. (\ref{theta}) becomes zero and $\theta_1^{T}$ in the magneto-optical metamaterial can be enhanced by almost an order \cite{met1}.
 As for the Faraday rotation with time dependent dielectric permittivity tensor, where $g=g_o \cos(\Omega t)$ and $\Omega$ is the angular frequency of the gyrotropic vector, it can be shown that the time dependent Faraday rotational angle, besides the standard term (\ref{theta}), contains an extra term which is proportional to time $t$ and $\frac{\omega}{\Omega}$ which increases faster than the stationary term and becomes dominant provided that $t\omega>1$ \cite{1995}.
 
%citation was \cite{GG2016}

\subsection{Real part of FR in RHM/LHM: Transmission}

Let us first consider the FR for transmission from a slab. Since the Faraday effect is typically very small the effective incident indices of refraction and impedance for the two circular polarizations in the first order of g can be presented in the form
$$n_{\pm}=\sqrt{\epsilon_{\pm}\mu}\approx n\pm\frac{1}{2}\frac{gn}{|\epsilon|},$$
$$z_{\pm}=\sqrt{\mu/\epsilon_{\pm}}\approx z\mp\frac{1}{2}\frac{gz}{|\epsilon|},$$
where $n$ (refractive index of a homogeneous material) and $z$ (impedance of a homogeneous material) are calculated when gyration vector $\vec g$ is zero.
Note, that by replacing $n\rightarrow -n$ we can use the above expressions for LHM.

\begin{figure}
\begin{center}
\includegraphics[width=8.4 cm]{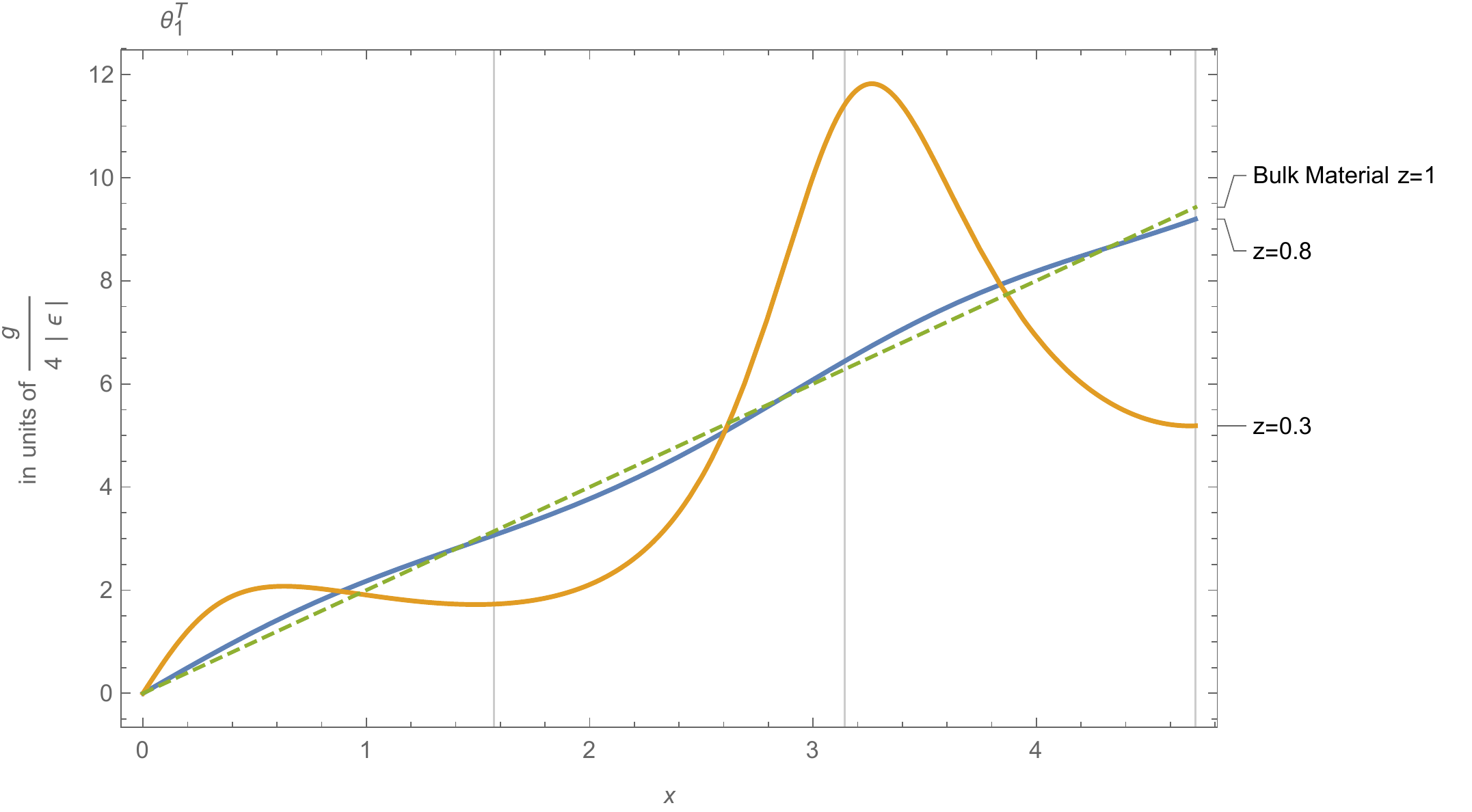}
\caption{Faraday Rotation angle $\theta_1^{T}$ as a function of $x=\omega nL/c$ for $z=0.3$ and $z=0.8$.}
\label{Real Transmission}
\end{center}
\end{figure}

One can simplify the analysis of $\theta_1^{T}$ and $\theta_2^{T}$ by expanding $\psi_{\pm}$ around the $n$ and $z$ of the slab in the absence of the magnetic field $\vec B$. Then the Taylor series of $T^{1/2}_{\pm}$ and $\psi_{\pm}$ in the neighborhood of $n$ \& $z$ becomes:
\begin{equation}
T^{1/2}_{\pm}=T^{1/2}(n,z)\pm \frac{1}{2}\frac{gn}{\epsilon}\frac{\partial T^{1/2}}{\partial n}\mp\frac{1}{2}\frac{gz}{\epsilon}\frac{\partial T^{1/2}}{\partial z}\label{T1},
\end{equation}
\begin{equation}
\psi_{\pm}=\psi(n,z)\pm\frac{1}{2}\frac{gn}{|\epsilon|}\frac{\partial \psi}{\partial n}\mp\frac{1}{2}\frac{gz}{|\epsilon|}\frac{\partial \psi}{\partial z}.
\end{equation}
Hence,
\begin{equation}
\theta_1^{T}=\frac{\psi_{+}-\psi_{-}}{2}=\frac{1}{2}\frac{gn}{|\epsilon|}\frac{\partial\psi}{\partial n}-\frac{1}{2}\frac{gz}{|\epsilon|}\frac{\partial \psi}{\partial z}=
\frac{1}{2}\frac{g}{|\epsilon|}\bigg(n\frac{\partial \psi}{\partial n}-z\frac{\partial \psi}{\partial z}\bigg).\label{1a}
\end{equation}
Evaluating the derivatives $\frac{\partial\psi}{\partial n}$ and $\frac{\partial\psi}{\partial z}$ at $\vec B =0$ from Eq.(\ref{p}) and substituting these expressions into Eq. (\ref{1a}) where, for convenience, we have introduced the new parameter $x=\omega nL/c$, we get
\begin{equation}
\theta_1^{T}=\frac{g}{4|\epsilon|z}\frac{x(z^2+1)+(1-z^2)\sin x\cos x}
{1+\frac{1}{4}\bigg(z-\frac{1}{z}\bigg)^2\sin^2 x},\label{2}
\end{equation}
Eq. (\ref{2}) is a general expression and valid for any continuous material with arbitrary parameters $L$, $n$ and $z$. As expected, $\theta_1^{T}$ is odd in $n$, where in LHM it will change sign of $n$. Below we analyze a few of the limits for these parameters:

When $L$ tends to zero ($kL\ll 1$) , the above equation reduces to
\begin{equation}
\theta_1^{T}\approx \frac{g}{2\epsilon z}x\equiv\frac{g\omega L}{2c}\frac{\epsilon}{|\epsilon|},
\end{equation}
which coincides with the RHM ($\epsilon >0$) thin-film result of Ref. \cite{zh}.

If $z=1$, i.e. when light propagates in a homogenous medium, we get
\begin{equation}
\theta_1^{T}=\frac{g}{2|\epsilon|}x\equiv \frac{g\omega L}{2c|\epsilon|}\sqrt{{\mu\epsilon}},\label{3}
\end{equation}
which coincides with the result of Refs. \cite{LL,zh} in the thick film limit where $kL\gg 1$, if $\mu=1$ for RHM (the range of all optical frequencies).
At the points $x_0=\frac{z^2+1}{z^2-1}\tan x_0$, where the ellipticity is zero when $\theta_2^{T}=0$, as was mentioned previously, and we get for the real part of FR
\begin{equation}
\theta_1^{T}=\frac{gz}{|\epsilon|(z^2+1)}x_0
\end{equation}
In Fig. 1, we show the FR angle of transmission vs $x=\omega nL/c$ for RHM, for three different values of surface impedance, using Eq. (\ref{2}): $z=0.3$, $z=0.8$, and bulk material with no reflections where $z=1$ (dashed line). The angle steadily increases and oscillates around the line $ \theta_1^{T}= 2x$ (Where $x$ is in units $\frac{g}{4|\epsilon|}$ ) with certain periodicity of $\pi$ or on the scale $L\sim k^{-1}$. The oscillations in $\theta_1^{T}$ are due to interference effects in the plane-parallel slab and the amplitude of the oscillating part depends on $x$. At $x_l={\pi}(l+1/2)$ we have for FR angle $\theta_1^{T}=\frac{gz}{|\epsilon|}\frac{x_l}{z^2+1}$ and for $x_l=\pi l$ ($l=1,2 ...$) $\theta_1^{T}=\frac{g}{4|\epsilon|}\frac{z^2+1}{z}x_l$. We were not able to find a closed-form solution analytically for the maximum of $\theta_1^{T}$, and Eq. (\ref{2}) and could not calculate the maximum increase of FR angle. However, for the estimated increase we used points $x_l=\pi l$, because the maximum value of $\theta_1^{T}$ for each period of oscillation is located very close to that point (see Fig. 1 where the vertical grid line appears). Then the ratio of $\theta_1^{T}$ at $x_l=\pi l$ to the $\theta_1^{T}$ in a homogeneous media, Eq. (\ref{3}), reads $(z^2+1)/2z \geq 1$. For materials with relative impedance $\sim 0.3$ (semiconductor with zero extinction coefficient in the near or mid infrared range like tellurium or aluminum gallium arsenide) the ratio is almost 2.  In otherwords, multiple reflections increase the overall time the light spends within the system showing an increase in Faraday rotation \citep{1995}.  A similar increase of Faraday rotation was also found in \cite{SPIE,met1}.
 However, for the composite system (dielectric with metamaterials or super lattice systems) the effective $\epsilon$ can be reduced up to $10^{-2}$ and the ratio can thus be increased by an order or greater.

\subsection{Imaginary part of FR in RHM/LHM: Transmission}

Expanding $T^{1/2}_{\pm}$ around the $n$ and $z$ of the slab in the absence of the magnetic field $\vec B$ (see Eq. (\ref{T1})) and using the Taylor series for $\ln (1 + x)$ centered at $0$ we can similarly
derive the expression for the ${\theta_2^{T}}=\frac{1}{2}\ln\frac{T^{1/2}_{+}}{T^{1/2}_{-}}$ for the imaginary portion of Faraday rotation as
\begin{equation}
\theta_2^{T}=\frac{g}{8|\epsilon|z^2}\frac{(1-z^2)\sin{x}\bigg[(z^2+1)\sin {x}+x(1-z^2)\cos{x}\bigg]}{1+\frac{1}{4}\bigg(z-\frac{1}{z}\bigg)^2\sin^2{x}}. \label{02}
\end{equation}
This is again a general expression and valid for the arbitrary parameters $L$, $n$ and $z$. As expected, $\theta^{T}_2$ is even in $n$, and $\theta^{T}_2\rightarrow 0$ when $L$ tends to zero. As it was previously mentioned, $\theta^{T}_2$ becomes 0 at z=1 (no boundaries), at $x=\pi l$ (complete transmission) and at $x_0=\frac{z^2+1}{z^2-1}\tan x_0.$ In the two former cases the coefficient of transmission $T$ becomes 1 when an external magnetic field $\vec B$ is zero. The third case is very different: The transmission coefficient is not 1 and $T\rightarrow 4z^2/(z^2+1)^2$ as $x_0$ tends $\infty$. This saturated value corresponds exactly to one-quarter wavelength.

Note, that in the limit of a small magnetic field $\vec B$, the expression for $\frac{b}{a}$, Eq. ({\ref{ration}}), coincides with $Im \theta^{T}$, that is with Eq. ({\ref{02}).

Fig. 2 shows for $z=0.3$ (solid) and $z=0.8$ (dashed) the imaginary angles of the FR, Eq. ({\ref{02}), for a RHM ($n > 0$) versus $x$. $\theta_2^{T}$ in the interval $[0,\pi]$ increases with $x$, reaches a peak value and then drops to become minimum at some point. This pattern repeats as $x$ increases.

\begin{figure}
\begin{center}

\includegraphics[width=8.4 cm]{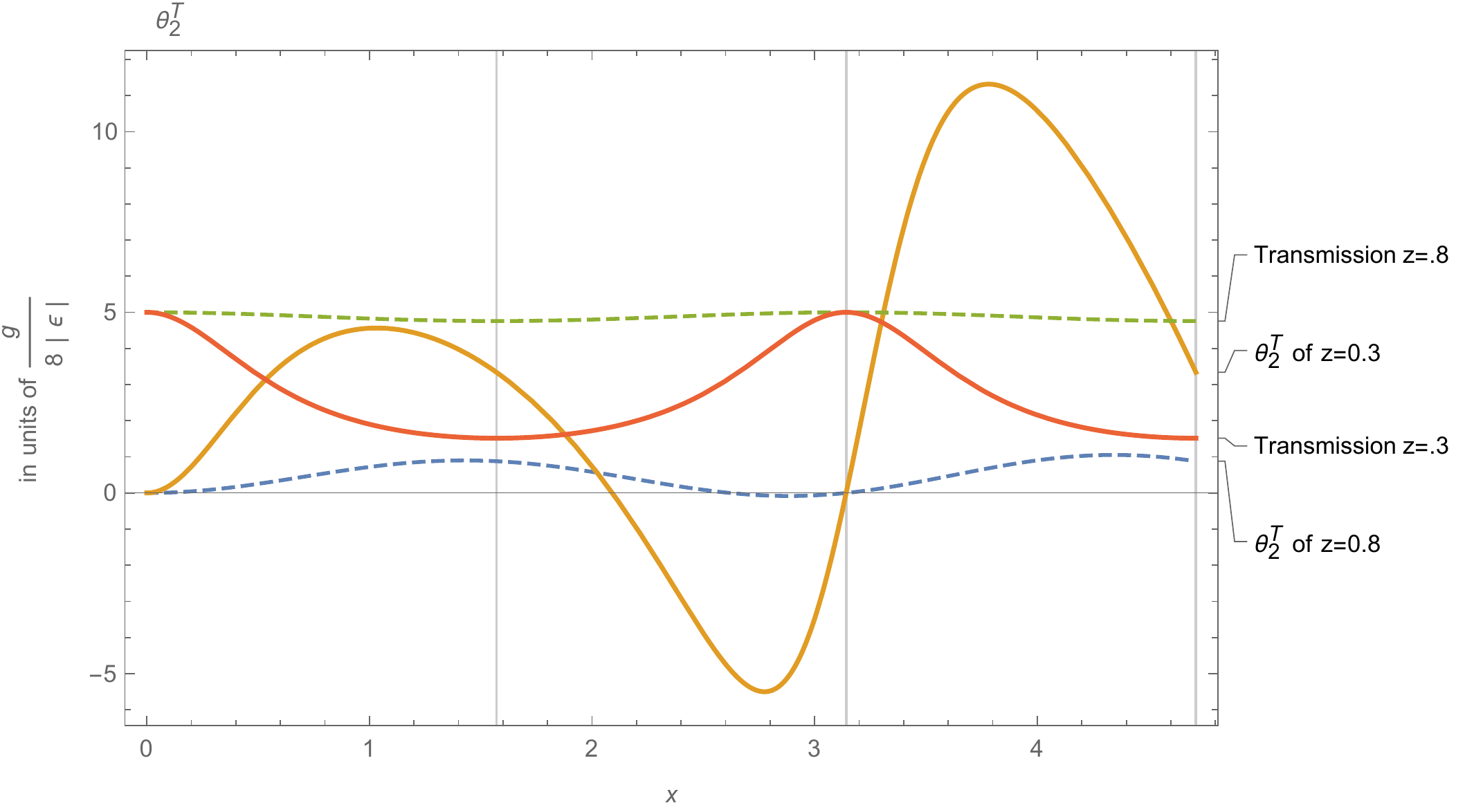}
\caption{Faraday Rotation angle $\theta_2^{T}$ as a function of $x=\omega nL/c$ for z=0.3 and z=0.8.  Transmission is multiplied by 5 to more easily see the relationship between the Faraday ellipticity angle and corresponding transmission value, where $T=1$ is the norm.}
\label{Imaginary Transmission}

\end{center}
\end{figure}

\section{Real and Imaginary parts of KR in RHM/LHM: Reflection}

When  linearly  polarized  light is  reflected  from  the  surface  of  a  magnetized  material,  the  direction  of  polarization  is  changed and  the  light  is  elliptically  polarized.  This is the Kerr effect and it is very  similar  to  the  Faraday  effect except  that  the  Kerr  effect  refers to the  reflection  and  the  Faraday  effect refers to the transmission.

Before entering into a more detailed analysis of the complex Kerr effect, let us note that if we ignore
the losses then there are some useful results which relate the $\theta^{T}$ and $\theta^{R}$ which follow already from the general expressions of the scattering matrix elements in terms of the transmission and reflection probabilities and the scattering phases $\psi$ and $\psi\pm \psi_a$. Here $\psi$ is the total phase accumulated in a transmission event and $\psi\pm \psi_a$ are the phases accumulated by a particle which is incident from either face of the material (left or right) which is reflected. The scattering-matrix elements can be written in the form
\[
S=\left( \begin{array}{cc}
r & t \\
t & r^{\prime}
\end{array}
\right)=\left( \begin{array}{cc}
-i\sqrt{R}\exp{i(\psi+\psi_a)} & \sqrt{T}\exp{i(\psi)}  \\
\sqrt{T}\exp{i(\psi)}  & -i\sqrt{R}\exp{i(\psi-\psi_a)}
\end{array}
\right)\]

For a spatially symmetric barrier the phase asymmetry $\psi_a$ vanishes and one has additionally $r=r^{\prime}.$

And it is clear that for any symmetric structure with no material loses, including the slab we are discussing,
$$\theta^{R}_1=\theta^{T}_1. $$

%\subsection*{Imaginary part of FR in right-handed materials: Reflection}
Whereas previously it had been desribed that $t_{\pm}$ was the transmission amplitude of the wave, we shall now describe $r_{\pm}$ as the reflection amplitude. It can be shown for the slab that the reflection amplitude is given by \cite{LL},
\begin{equation}
r_{\pm}=-i\frac{t_{\pm}}{2}(z_{\pm}-\frac{1}{z_{\pm}})\sin(n_{\pm}\omega L/c)
\end{equation}
where $t_{\pm}$ is defined by Eq. (\ref{ta}).

Using a similar expression for Kerr effects complex reflection angle
\begin{equation}
\theta^{R} =-i\ln\frac{r_{+}}{r_{-}}=\theta^{R}_1+i\theta^{R}_2, \label{FR2}
\end{equation}
where,
\begin{equation}
\theta^{R}_2=\theta^{T}_2-\frac{g}{2\epsilon}\bigg(\frac{(nkL)\cos(nkL)}{\sin(nkL)}+\frac{z^2+1}{1-z^2}\bigg), \label{KR2}
\end{equation}
where $\theta^{T}_2$ is defined by Eq.(\ref{02}).

When $L$ tends to zero (the thin film approximation), the above expression reduces to $$\theta^{R}_2\approx\frac{\epsilon}{|\epsilon|}\frac{g}{\epsilon-\mu}.$$ where the unit vector of epsilon has a sign change from RHM to LHM.  

As seen from the above expression, $Im \theta^{T}$ is proportional to the extremely small parameter $g$ and in RHM, where $\mu\ll \epsilon$, it is too difficult to measure $\theta^{R}$. However, the situation is very different for LHM, where $\mu$ and $\epsilon$ can be of the same order of magnitude for some frequency range (for example $\mu$ and $\epsilon$ for NZE metamaterials). For these frequencies it can be verified experimentally that a narrow resonantly enhanced reflection angle can be found for the Kerr effect.

Fig. 3 shows the imaginary angle of the KR, Eq. ({\ref{KR2}), for two different RHM of different surface impedance $z$ versus $x$ . $\theta_2^{R}$ at $x_l=\pi l$ shows a discontinuity.  We also note that zeroes for both $\theta_2^T$ and $\theta_2^R$ coincide and are the solutions to the transcendental equation $x_0=\frac{z^2+1}{z^2-1}\tan x_0$.  At this point there is linearly polarized light for both the reflected and transmitted light.  
\begin{figure}
\begin{center}
\includegraphics[width=8.4cm]{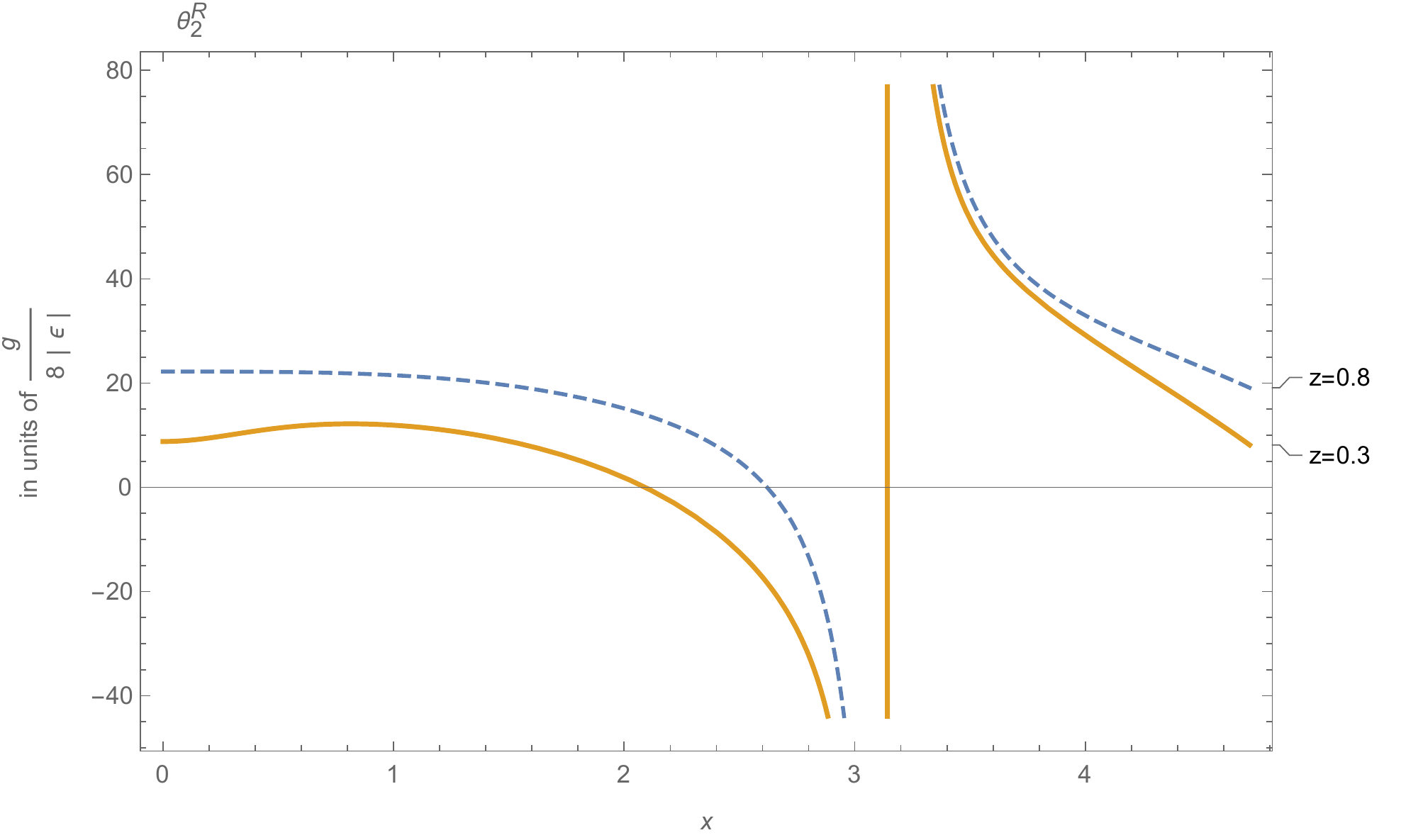}
\caption{Kerr Rotation angle $\theta_2^{R}$ as a function of $x=\omega nL/c$ for $z=0.3$ and $z=0.8$.}
\end{center}
\end{figure}

\section{Summary}

We study the Faraday and Kerr rotations of light with angular frequency $\omega$ passing through a RHM/LHM  film with thickness L while taking into account the multiple reflections from the boundaries.  The descriptions of the real portions as the linear angle of rotation and imaginary portions as the ellipticity of the rotation allow us to separate the two distinct phenomena and visualize their maximums and effects within different kinds of mediums.  We found that the rotation and ellipticity of the transmitted or reflected light has shown that the real parts of the complex angle of the Faraday and Kerr effects are odd functions with respect to the refractive index n. As well, the imaginary portion of the angle is an even function of n. These odd and even functions are not just the properties of a thin film,  but apply just as well to the case of any system of arbitrary length.

For a spatially symmetric film with no material loses the real portion of Faraday and Kerr rotations are equal for RHM and LHM. In the limit of an ultra thin LHM film under specific circumstances a large resonant enhancement of the reflected KR angle could be experimentally obtained.  From this it has been shown that with multiple reflections within the medium that the maximums of the real portions of the Faraday and Kerr effects do not coincide with simultaneously zero imaginary portions (figure 2).  This means that the maximums of both Faraday and Kerr rotations occur only when the light has some ellipticity, or with non-zero imaginary portions.  Taking into account these multiple reflections also shows the resonant enhancement that is now possible with LHM such as the super lattice system, and opens the field of optics to new compositions of materials that can greatly enhance these rotations by an order or more.

%\bibliographystyle{unsrt}

%\bibliography{AJP092216}

\end{document}